\documentstyle[12pt]{article}
\topmargin--0cm
\oddsidemargin--1mm
\begin{document}
\def\<{\langle}
\def\>{\rangle}
\newcommand{\pl}{\partial}
\newcommand{\be}{\begin{equation}}
\newcommand{\ee}{\end{equation}}
\newcommand{\ba}{\begin{eqnarray}}
\newcommand{\ea}{\end{eqnarray}}
\def\R{\relax{\rm I\kern-.18em R}}
\def\1{\relax{\rm 1\kern-.27em I}}
\newcommand{\mb}[1]{\mbox{\boldmath${\bf #1}$}}
\newcommand{\Z}{Z\!\!\! Z}
\newcommand{\ph}{PS_{ph}}
\begin{center}
{\LARGE The Monopole Dominance in QCD }

\vskip 0.5cm
{Sergei V. SHABANOV}\footnote{\noindent
Alexander von Humboldt fellow;\\
on leave from Laboratory of Theoretical
Physics, JINR, Dubna, Russia.}

\vskip 0.5cm
{\em Institute for Theoretical Physics, FU-Berlin, 
Arnimallee 14, D-14195, Berlin, Germany}
\end{center}

\begin{abstract}
A projection (gauge) independent  formulation of the
monopole dominance, discovered in lattice QCD for
the maximal abelian projection,
is given. A new dynamical abelian projection
of continuum QCD, which does not rely on any
 explicit gauge condition imposed on gauge fields,
is proposed. Under the assumption that the results
of numerical simulations hold in the continuum limit,
the monopole dominance is proved for
the dynamical abelian projection. The latter enables
us to develop an effective scalar field theory for
dominant (monopole) configurations of gauge fields.
The approach is manifestly gauge and Lorentz invariant.
\end{abstract}

\subsubsection*{1. The Coulomb problem in QCD}

To calculate an interaction energy $V(R)$ of two static
color sources separated by a distance $R$ in non-abelian
gauge theory, one has to analyze the behavior of the 
Wilson loop expectation value in the large Euclidean time
limit
\begin{equation}
\< W_R\>=\< 1\>^{-1} 
\int\limits_{}^{}{\cal D}A_\mu e^{-S_{YM}}{\rm tr}\,
Pe^{ig\oint_{C_R}dx_\mu A_\mu}
\rightarrow e^{-TV(R)}\ ,
\label{3}
\end{equation}
as $T\rightarrow \infty$. The integral (\ref{3}) is taken over
configurations satisfying periodic boundary conditions
$A_\mu (\mb{x},0) =A_\mu(\mb{x}, T)$, and $S_{YM}$ is the
Euclidean Yang-Mills action for a finite Euclidean time $T$;
the exponential $e^{-S_{YM}}$ serves as the probability distribution
in (\ref{3}).

In electrodynamics fluctuations of electromagnetic fields are gaussian
and described by an infinite set of harmonic oscillators
driven by an external force. Doing the integral (\ref{3})
and taking the limit, one finds that
$V(R)\sim R^{-1}$, i.e. the Coulomb law. In the case of Yang-Mills theory,
the integral is not gaussian. The theory describes an infinite set of
coupled anharmonic oscillators driven by an external force.
Therefore, some approximate method to evaluate the integral (\ref{3})
should be developed.
It is well known that the perturbation theory for the integral (\ref{3})
leads to the Coulomb law at short distances, which reflects the asymptotic
freedom in QCD, whereas for large distances, the perturbation expansion breaks
down and fails to reproduce the expected linearly raising potential
\begin{equation}
V(R)=V_0 + \sigma R+O(R^{-1})\ 
\label{6}
\end{equation}
that provides the quark confinement.

A difficulty to develop an approximate
approach is due to the lack of understanding what
configurations of gauge fields give a main contribution to the integral
(\ref{3}) and, hence, are responsible for the QCD confinement mechanism.
A priori there is no clue for how these configuration may look like.
Fortunately, recent results of numerical simulations of the lattice QCD
show that such configurations do exist, and they look
like Dirac magnetic monopoles
when taken in a specific gauge called the maximal abelian gauge.
Therefore, it is natural to attempt to construct an effective theory for
their dynamics. The present letter is devoted to this problem.

\subsubsection*{2. Monopoles in abelian projections}

The idea to associate the dominant configurations with magnetic
monopoles attracts much attention because it gives a transparent
physical picture for the confinement mechanism where the QCD vacuum
is assumed to be a dual superconductor formed by condensed
monopole-antimonopole pairs \cite{thooft1}.
Magnetic monopoles, as classical soliton-like excitations, 
are hard to introduce in the non-abelian gauge
theory with a simply connected compact group (like, e.q. SU(3))
without a Higgs mechanism, i.e. when the gauge symmetry remains unbroken.
A way out
was suggested by 't Hooft who proposed a partial gauge fixing,
$\chi(A)=0$, to
restrict the gauge group $G$ to its maximal abelian subgroup $G_H$
\cite{thooft2}.
When lifted onto the gauge fixing surface $\chi(A)=0$ by a
suitable gauge transformation, gauge potentials
appear to have topological singularities
occurring through the lift (projection). 
For the unitary groups $G=SU(n)$
these singularities can be viewed as magnetic monopoles relative to the
unbroken electromagnetic group $G_H=(U(1))^{n-1}$.

To see how the topological singularities occur in an
abelian projection, let us pick up a local
gluon operator $\Gamma (A)$ that transforms according to the adjoint
representation
\begin{eqnarray}
\Gamma (A)&\rightarrow & \Gamma (A^\Omega)
=\Omega \Gamma (A) \Omega ^\dagger\ ,
\label{7} \\
 A_\mu&\rightarrow &A_\mu ^\Omega
=\Omega A_\mu \Omega ^\dagger +ig^{-1}\Omega \pl_\mu
\Omega ^\dagger\ .
\label{8}
\end{eqnarray}
The gauge condition breaking $G$ to $G_H$ is
\begin{equation}
\chi(A)=\Gamma ^{off}(A)=0\ ,
\label{9}
\end{equation}
where "off" stands for the off-diagonal elements of $\Gamma$ in a matrix
representation. For example, one can take $\Gamma (A)=F_{12}(A)$. Clearly,
the condition (\ref{9}) is invariant relative to
the abelian transformations from $G_H$ and,
hence, does not break $G_H$. Given a configuration $A_\mu $,
consider a gauge group element $\Omega _\Gamma$ such that
the element
\begin{equation}
\Gamma _H =\Omega _\Gamma \Gamma (A)\Omega ^\dagger_\Gamma
\label{10}
\end{equation}
belongs to the Cartan subalgebra (i.e. diagonal in the matrix
representation).
Then the potential
\begin{equation}
A^{\Omega_\Gamma}_\mu =\Omega _\Gamma A_\mu \Omega ^\dagger_\Gamma +ig^{-1}
\Omega _\Gamma \pl _\mu \Omega ^\dagger_\Gamma
\label{11}
\end{equation}
satisfies the gauge condition (\ref{9}). In other words,
the gauge transformation (\ref{11}) lifts the configuration $A_\mu $
onto the gauge fixing surface
(\ref{9}); the function $\chi(A^{\Omega_\Gamma})$ identically vanishes
for any configuration $A_\mu$ and the corresponding group element
$\Omega_\Gamma$.

Doing this abelian projection for all configurations $A_\mu$, we arrive
at the abelian gauge theory where diagonal (Cartan) components,
denoted below as $C_\mu^\Gamma$,
of the projected potentials (\ref{11})
play the role of the Maxwell fields,
while the off-diagonal components are charged fields. We set
\begin{equation}
A^{\Omega_\Gamma} _\mu =C^\Gamma _\mu +W^\Gamma_\mu\ .
\label{12}
\end{equation}
Under residual gauge transformations, they behave as
\begin{equation}
C_\mu^\Gamma \rightarrow C_\mu^\Gamma +\pl _\mu \omega \ ;\ \ \
 W_\mu^\Gamma \rightarrow 
 e^{ig\omega}W_\mu^\Gamma e^{-ig\omega}\ ,\ \ \ \ e^{-ig\omega}\in
 G_H\ .
\label{14}
\end{equation}

The abelian theory thus obtained is not the usual one. Not for every
configuration $A_\mu$, the projective group element $\Omega_\Gamma(A)$
determined by (\ref{10}) is a well defined function in spacetime.
Let us take, for example, the group SU(2), then $\Gamma = \Gamma_a
\tau_a$, $\tau_a$ are the Pauli matrices, is a 2$\times$2 traceless
hermitian matrix with elements being functions of spacetime. The
group element $\Omega_\Gamma = \Omega_\Gamma (x)$ is well defined
everywhere in spacetime, but the points where $\Gamma_a(x)$ vanish.
Three equations $\Gamma_a(x)=0$ for four spacetime coordinates
 determine a
worldline $x_\mu =x_\mu (\tau)$ (or a set of worldlines).
Through the projection transformation (\ref{11}) the singularities of
$\Omega_\Gamma$
are transferred to the potentials $A_\mu^{\Omega_\Gamma}$.
Singularities of the Maxwell fields appear to be Dirac magnetic
monopoles so that the equations
$\Gamma _a(x)=0$ determine their worldlines.

In the SU(2) case, the Maxwell field has only one component
\begin{equation}
C_\mu^\Gamma ={\textstyle{\frac 12}} {\rm tr}\,
\tau_3\left( \Omega_\Gamma A_\mu\Omega_\Gamma^\dagger
+ ig^{-1}\Omega_\Gamma \pl_\mu\Omega_\Gamma^\dagger\right)\ .
\label{14a}
\end{equation}
Let a configuration $A_\mu$ be such that $\Gamma_a(\mb{x}_m)=0$
at some isolated spatial point $\mb{x}=\mb{x}_m$ (time is fixed).
Imagine a sphere $\Sigma_m$ centered at $\mb{x}_m$ and a closed
contour $L_s$ on it. The magnetic field flux through $L_s$ is
\begin{equation}
\Phi_{L_s} = \oint_{L_s} (\mb{C}^\Gamma, d\mb{l}) \ .
\label{14b}
\end{equation}
Consider the limit when the contour $L_s$ shrinks to some
point $\mb{x}_s$ on the sphere. If $\mb{C}^\Gamma$ is regular
everywhere on the sphere, the flux vanishes, $\Phi_{L_s\rightarrow 0}
=0$, for any $\mb{x}_s$. Since $\Omega_\Gamma$ is regular on the
sphere, the first term in (\ref{14a}) is regular too and gives no
finite contribution to the flux (\ref{14b}) in the limit
$L_s\rightarrow 0$. In contrast, the second term can contain
a total derivative of angular (cyclic) variables parametrizing
the group element $\Omega_\Gamma$ in the reference frame
originated at $\mb{x}_m$. If this happens, the flux
$\Phi_{L_s\rightarrow 0} $  does not vanish at some points
$\mb{x}_s$ on the sphere $\Sigma_m$.  The latter means that
the magnetic field $\mb{B}^\Gamma= \mb{\nabla}\times \mb{C}^\Gamma$ has
a string-like singularity which carries a finite magnetic flux
and, therefore, can be associated with the Dirac string of
a magnetic monopole that pierces $L_s$. An explicit example
of a monopole-carrying group element $\Omega_\Gamma$ is given
in section 4.

The origin of this topological singularity can be understood
in the following way. On the sphere $\Sigma_m$ we define
a compact field $e_\Gamma = \Omega^\dagger_\Gamma
\tau_3\Omega_\Gamma$. Clearly,
it is regular on the sphere and, hence, determines a map
of the sphere $\Sigma_m $ onto the sphere 
${\rm tr}\, e_\Gamma^2 = 2$ in the isotopic space. 
This map is characterized by the Poincare-Hopf index, being
the number of times one sphere is wrapped about the other,
\begin{equation}
q_\Gamma(x_m) =(32\pi i)^{-1}\oint_{\Sigma_m}d\sigma_j
\epsilon_{jkn}{\rm tr}\, (e_\Gamma[\pl_k e_\Gamma, \pl_n e_\Gamma])\ .
\label{14c}
\end{equation}
Making use of the Stokes theorem we write (\ref{14b}) as a surface
integral
\begin{equation}
\Phi_{L_s}= - \int_{\Sigma^s_m}d\sigma_j \epsilon_{jkn}\pl_kC^\Gamma_n
=- \int_{\Sigma^s_m}(d\mb{\sigma},\mb{B}^\Gamma)\, ,
\label{14d}
\end{equation}
where $\Sigma^s_m$ is the sphere $\Sigma_m$ with a small hole cut out
by the contour $L_s$. In the limit $L_s\rightarrow 0$, the surface
$\Sigma^s_m$ turns into $\Sigma_m$ without the point $\mb{x}_s$.
Substituting the singular part of (\ref{14a}),
\begin{equation}
\bar{C}^\Gamma_n =- (2ig)^{-1}{\rm tr}\, (\tau_3\Omega_\Gamma\pl_n
\Omega_\Gamma^\dagger) =
(2ig)^{-1}{\rm tr}\, (e_\Gamma\Omega_\Gamma^\dagger\pl_n \Omega_\Gamma)\ ,
\label{14e}
\end{equation}
in (\ref{14d}),
we find after simple algebraic transformations
\begin{eqnarray}
\Phi_{L_s\rightarrow 0} &=& {\textstyle{\frac{1}{16 ig}}}
\oint_{\Sigma_m^s}d\sigma_i\varepsilon_{ijk}
{\rm tr}\, \left(e_\Gamma [\pl_j e_\Gamma,\pl_k e_\Gamma]\right) +
{\textstyle{\frac{1}{2ig}}} 
\oint_{\Sigma_m^s} d\sigma_i\varepsilon_{ijk}
{\rm tr}\, \left(\tau_3\Omega_\Gamma\pl_j\pl_k\Omega_\Gamma^\dagger
\right)
\label{14dd}\\
&=&2\pi g^{-1}q_\Gamma(x_m)\equiv 4\pi g_m\ .
\label{14f}
\end{eqnarray}
At the singular point $\mb{x}_s$, we have $[\pl_j,\pl_k]\Omega_\Gamma
\neq 0$ so that the last term in (\ref{14dd}) does not contribute
(the Dirac string pierces the sphere $\Sigma$ at this point).
Therefore
the magnetic charge of the Dirac monopole located at $\mb{x}=\mb{x}_m$
is equal to $g_m =-q_\Gamma/2g$.

Thus, after the abelian projection we obtain  electrodynamics
with Dirac magnetic monopoles that are
topological singularities in the abelian part of the projected gauge
fields.

The construction can be extended to any abelian projection
determined by a gauge condition $\chi(A) =0$ that breaks
the gauge group to its maximal abelian subgroup. For every
configuration $A_\mu$ we determine a gauge group element
$\Omega_\chi$ by the equation
\begin{equation}
\chi(A^{\Omega_\chi}) =0 \ .
\label{chi}
\end{equation}
If $\lambda^\alpha$, $\alpha =1,2,..., {\rm rank}\, G$,
is an orthonormal
basis in the Cartan subalgebra, then the scalar fields,
that characterize topological singularities in
the Maxwell fields $C_\mu^\chi =\lambda^\alpha
{\rm tr}\, (\lambda^\alpha A_\mu^{\Omega_\chi})$, are defined
as $e_\chi^\alpha = \Omega_\chi^\dagger \lambda^\alpha\Omega_\chi$.
They determine a map of the sphere $\Sigma_m$ to the coset space
$G/G_H$.
The homotopy group of this mapping is
$\Pi_2(G/G_H) = \Pi_1(G_H) = Z\!\!\!Z^{{\rm rank}\ G}$ (here
$G=SU(n)$). Therefore
the monopole distribution and their
charges are still determined by ${\rm rank}\ G$ relations
(\ref{14c}).
Every configuration $A_\mu$ is thus
associated with a certain set integer-valued functions $q_\chi^\alpha(x)$.

So, the abelian projection
$\chi$ divides the space $[A]$ of all configurations $A_\mu$
into two subspaces characterized
by $q_\chi \neq 0$ and by $q_\chi =0$. Upon the abelian
projection $\chi$, the abelian components of
configurations from the former ("monopole") subspace turn into the Dirac
magnetic monopole potentials with 
the magnetic charge distribution $q_\chi$,
while configurations from the other subspace give rise to no monopole
after the projection. We remark that, following
a tradition in lattice gauge theories we call the subspace
with $q_\chi\neq 0$ the "monopole" subspace,
but we put the quotation marks in order to emphasize the fact
that {\em before} the $\chi$-projection, configurations from
this subspace are no Dirac monopoles.
The above two subspaces in $[A]$ 
are defined up to {\em regular}
gauge transformations. Note that the magnetic charge distribution
(\ref{14c})  is invariant under transformations
\begin{equation}
\Omega_\chi \rightarrow \Omega_0\Omega_\chi\ ,
\label{chi2}
\end{equation}
where $\Omega_0$ is regular in spacetime. It is also important to realize
that the above definition of the "monopole" subspace is projection
(or {\em gauge}) dependent. Different choices of $\chi$
lead, in general, to different
``monopole'' subspaces in $[A]$. In next section we give a gauge invariant
(projection independent) definition
of the "monopole" subspace {\em selected} by abelian projections.

\subsubsection*{3. Gauge invariance and the monopole dominance}

In lattice gauge theory, the integral (\ref{3}) can be done numerically.
As expected, the potential extracted from the lattice simulations has the
form (\ref{6}), and the coefficient $\sigma$ (called
the QCD string tension) is known. Now one can perform the abelian
projection on the lattice \cite{kronfeld}, 
i.e. the above described procedure
of dividing the space $[A]$ of all configurations
$A_\mu$ into the "monopole" and "non-monopole" subspaces
can be performed numerically for a given projection $\chi$.
To understand how big a contribution
of the monopole configurations to the  QCD string tension, one has just to
restrict in (\ref{3}) the sum over configurations by the
"monopole" subspace and evaluate the string tension.
It turns out that for
the so called maximal abelian projection \cite{thooft2,kronfeld}
\begin{equation}
\chi_{ma}(A)=\pl _\mu A^{off}_\mu +ig [A^H_\mu ,A^{off}_\mu ]=0\ ,
\label{15}
\end{equation}
where $A_\mu =A^H_\mu +A^{off}_\mu$, and $A^H_\mu$ are Cartan (diagonal)
components of $A_\mu$,
the difference between an exact string tension and the string tension
evaluated on the monopole configurations is only eight
per cent \cite{stack} (see also recent simulations \cite{bali}). 
This phenomenon
is known as the monopole dominance.

Based on the fact of the monopole dominance, one can conjecture
that the confinement is due to the condensation of monopole-antimonopole
pairs. However, such a
conclusion is not straightforward. Recall that this mechanism
is theoretically well understood only in gauge theories with the spontaneous
gauge symmetry breaking \cite{pol}, where monopoles are
solitons of classical equations of motion. In QCD the color local
symmetry remains unbroken and, therefore, the monopole-antimonopole
condensation in the
abelian projection, if it occurs, is not due to the usual Higgs mechanism.

It should be noted that the results of numerical simulations suggest only
that in the space of all configurations $[A]$ there is a relatively "small"
subset $[\bar{A}]$ that gives a dominant
contribution to the expectation value
of the Wilson loop. It seems also that
a major part of the dominant subset $[\bar{A}]$ can be {\em selected}
via the maximal abelian projection (\ref{15}).
A natural question arises from the results of numerical simulations.
Does there exist any gauge invariant formulation of
the monopole dominance? In fact, the monopole dominance has been
observed only in the maximal abelian projection and it is absent
in other projections studied on the lattice, although this does not
exclude the existence of some other abelian projections that
could exhibit the monopole dominance \cite{digiacomo}.
Below we answer this question.

Consider an abelian projection generated by a gauge condition $\chi(A)=0$.
Topological singularities can only occur in the projected configuration
$A_\mu^{\Omega_\chi}$ through singularities in $\Omega_\chi$ determined
by equation (\ref{chi}). Were the group elements $\Omega_\chi$  regular
for all configurations $A_\mu$, there would exist a global gauge condition
(in the mathematical language, a global cross-section in the space
of all connections $[A]$) determined by
$\chi(A)=0$ and $\pl_\mu C_\mu =0$, where $C_\mu$ is the abelian
component of $A_\mu$, which is impossible according
to the Singer theorem \cite{singer}.
Note that the gauge condition $\pl_\mu C_\mu=0$
breaks the residual abelian gauge group. Since Singer's arguments
do not apply to abelian gauge theories (the condition
$\pl_\mu C_\mu=0$ exists globally), we conclude that any abelian projection
exhibits topological singularities. In some sense, it is not an accident
that configurations having topological defects  {\em after} the projection
turn out to be dominant. These configurations are "pure non-abelian"
(homotopicaly nontrivial) and contain information about the gauge orbit
space topology in Yang-Mills theory.

Topological defects can generally occur in both the abelian $C_\mu^\chi$
and non-abelian $W_\mu^\chi$ components of projected
configurations, depending on the choice of $\chi$. To those occurring
in $C_\mu^\chi$ we shall refer as monopoles. One can construct an
abelian projection where no monopole singularities are possible, i.e.
all singularities occur in $W_\mu^\chi$. It was conjectured that
such singularities may also be associated with
dominant configurations \cite{polikarpov}, although
it has not been verified numerically yet. In what follows we consider only
monopole singularities. It is worth emphasizing again that the
dominant configurations do not depend on any gauge choice. The
gauge fixing is used only to {\em select} the dominant configurations
via topological defects occurring upon the abelian projection, and,
by now, there is no theoretical explanation of why
such a selection works in certain abelian projections (see section 5). 
In general,
there might exist some other ways to characterize them by topological
quantities different from the monopole distribution (\ref{14c}) (for
instance, in \cite{markum} a correlation between the instanton and
monopole topological numbers has been observed).

To specify the dominant "monopole" subset associated with a projection
$\chi$, we invoke another remarkable result of numerical simulations
known as the abelian dominance \cite{suzuki}.
It states that the so called abelian
string tension, the one that is calculated by averaging the Wilson
loop over only the abelian components $C_\mu^\chi$ of the projected
configurations $A_\mu^{\Omega_\chi}$, differs from the full QCD
string tension by approximately
eight per cent. Therefore the off-diagonal components
$W_\mu^\chi$ of the projected configurations are irrelevant for the
formation of the flux tube between static sources in QCD.
This has been
again verified in the maximal abelian projection and may not be the case
in another projection.  The dominant subset amongst the projected
configurations used to obtain the abelian string tension is wider than
the set of monopole configurations in $C_\mu^\chi$ because it
involves fluctuations of monopole-free Maxwell (or photon) fields.
The monopole dominance implies that contributions of photons to
the string tension  are
negligible as compared with that of Dirac monopoles \cite{stack}.

In the lattice simulations, the off-diagonal components
 $W_\mu^\chi$ of the projected configurations are set therefore 
to be zero.
It is not acceptable in the continuum case because the magnetic
field energy
of Dirac monopoles $\bar{C}_\mu^\chi$ (\ref{14e}) is infinite
as the monopoles are pointlike particles. Note that in the lattice
gauge theory the lattice spacing plays the role of the regularization
parameter at short distances.
Since the string tension is not sensitive to a particular
form of $W_\mu^\chi$, we can choose them to provide a core (a finite
size) for the Dirac monopoles. An explicit construction of the core
functions can be found in \cite{kogut}. With every monopole distribution
$q_\chi(x)\neq 0$ obtained in the abelian projection $\chi $
we associate the Dirac monopole
configuration $\bar{C}^\chi_\mu$ (\ref{14e}) and  a core function
$W_\mu^\chi =\bar{W}_\mu^\chi(\bar{C})$ corresponding to it (the Wu-Yang
monopole with a core).
In the projected theory, the dominant configurations are $\bar{C}_\mu^\chi
+ \bar{W}_\mu^\chi$. The color magnetic energy of these configurations
is finite, so is the action at finite temperature.

We define a subspace $[ ^\chi \bar{A}]$ in $[A]$ by the condition
that  any configuration $ ^\chi \bar{A}_\mu$ from
$[ ^\chi \bar{A}]$ becomes a Dirac monopole with a core
when projected on the gauge fixing surface $\chi = 0$, that is,
\begin{equation}
^\chi \bar{A}_\mu^{\Omega_\chi} = \bar{W}_\mu^\chi(\bar{C}^\chi)
+ \bar{C}_\mu^\chi\ .
\label{dc}
\end{equation}
The set $[q_\chi ]$ of all
possible monopole distributions $q_\chi$ occurring in an
abelian projection $\chi$ is isomorphic to the subspace $[ ^\chi \bar{A}]$
modulo {\em regular} gauge transformations. Note that due to
the invariance of $q_\chi(x)$ under the transformations (\ref{chi2}),
configurations $ ^\chi\bar{A}_\mu$ and  $ ^\chi\bar{A}_\mu^{\Omega_0}$
yield the same $q_\chi(x)$ upon the projection.

Consider a subspace $[\bar{A}]$ in $[A]$ that is a union of the spaces
$[ ^\chi\bar{A}]$ for all possible projections,
\begin{equation}
[\bar{A}]=\cup _{\chi}  [^\chi \bar{A}]\ .
\label{dc2}
\end{equation}
By construction, this subspace is {\em projection independent}. For any
configuration $\bar{A}_\mu$ from $[\bar{A}]$ there exists an abelian
projection gauge $\chi$ such that $\bar{A}^{\Omega _\chi}_\mu$ is the
vector potential of the form (\ref{dc}),
i.e. it describes a set of Dirac monopoles with cores.
The subspace (\ref{dc2}) modulo regular gauge transformations is
isomorphic to a set of all monopole distributions $[\bar{q}]$
that can be obtained in
all possible abelian projections of QCD
\begin{equation}
[\bar{q}] = \cup_\chi [q_\chi] \sim [\bar{A}]/[\Omega_0] \ .
\label{dc3}
\end{equation}

If the monopole dominance occurs, at least, in one abelian projection,
then the gauge invariant subspace $[\bar{A}]/[\Omega_0]$
dominates in the path integral (\ref{3}) because it is larger
than any of the "monopole" subspaces $[ ^\chi\bar{A}]$.  
At this point
we assume that the monopole dominance discovered for the maximal
abelian projection in lattice QCD survives the continuum limit
and, therefore, the integral (\ref{3}) is dominated by configurations
from $[\bar{A}]$. The dependence of the monopole dominance on
the abelian projection choice is easily understood.
The projection (or gauge) dependent "monopole" space $[ ^\chi\bar{A}]$
may or may not cover a major part of the projection independent
space (\ref{dc2}), depending on the luck in the abelian
projection choice. The maximal abelian projection
seems to be the "lucky" one.

Our next problem is to introduce a set of collective coordinates
in the projection independent "monopole" space $[\bar{A}]$.

\subsubsection*{4. Dynamical abelian projection}

Though we know that the "monopole" subspace associated
with the maximal abelian gauge covers a sufficiently large
part of the dominant set $[\bar{A}]$, technical difficulties
to find a parametrization of the corresponding subspace
$[ ^\chi \bar{A}]$ are overwhelming.
One has to solve equation (\ref{chi}) for the maximal abelian
gauge (\ref{15}) and a generic configuration 
$A_\mu$, which is a non-linear
differential equation for $\Omega_\chi$.
If a solution $\Omega_\chi(A)$
yields a non-zero monopole distribution (\ref{14c}),
then the corresponding configuration $A_\mu$ belongs to
the dominant subspace. In addition,
one has to bear in mind that equation (\ref{chi}) may have many
solutions (for a fixed $A_\mu$)
that are Gribov's copies of each other. In the lattice
QCD, the Gribov problem can be resolved numerically \cite{gribov},
though it does not seem to be of some relevance \cite{bali}.

In contrast to the maximal abelian projection, abelian projections
based on a local gluon operator $\Gamma$ in the
adjoint representation is technically simpler because the solving
of (\ref{chi}) implies a diagonalization of the matrix $\Gamma(A)$.
However there is neither theoretical nor numerical proof of the monopole
dominance in such abelian projections. They are also often in conflict
with the manifest Lorentz invariance.

To resolve the technical difficulties, we propose a new dynamical
abelian projection that does not rely on any explicit gauge condition
imposed on gauge field configurations. Its lattice version was suggested
and studied in \cite{lata}. The advantages of this projection are:
{\em (i)} monopole configurations are the only topological defects
in this projection, {\em (ii)} they are parametrized by a real scalar
field in the adjoint representation, {\em (iii)} the projection
is Lorentz invariant, and {\em (iv)} the Gribov problem is avoided.
Let us extend the lattice approach of \cite{lata} to the continuum case.

Consider the identity
\begin{equation}
1= \sqrt{\det (-D^2_\mu )}
\int{\cal D}\phi
e^{-\frac{1}{2}\int d^4x
{\rm tr} (D_\mu \phi)^2}\ ,
\label{17}
\end{equation}
where $D_\mu \phi =\pl _\mu \phi +ig [A_\mu ,\phi ]$, i.e. the scalar field
$\phi$ realizes the adjoint representation of the gauge group, and
substitute it in the path integral (\ref{3}) for the partition
function. As a result we obtain a theory where gauge
fields are coupled to an
auxiliary scalar field in the adjoint representation.
The idea is to use the auxiliary scalar field to make an abelian
projection.  The effective
action is invariant under the gauge transformations (\ref{8}) and
$
\phi \rightarrow \phi ^\Omega =\Omega \phi \Omega ^\dagger .
$
Instead of using a local gluon operator $\Gamma (A)$ to make an abelian
projection we set $\Gamma =\Gamma (\phi )=\phi$ and require
\begin{equation}
\Gamma ^{off}=\phi ^{off}=0\ .
\label{19}
\end{equation}
The gauge condition (\ref{19})
breaks the gauge group $G$ to its maximal abelian
subgroup $G_H$.

The projection is carried out as follows. Given a
configuration $\phi$, we find $\Omega _\phi $ such that $\Omega _\phi \phi
\Omega ^\dagger_\phi =h$ belongs to the Cartan subalgebra. The initial set of
field configurations $(\phi, A_\mu)$ is then lifted onto the gauge
condition surface (\ref{19}) by a gauge transformation with the group
element $\Omega _\phi$
\begin{equation}
(\phi ,A_\mu ) \rightarrow
(h, A^\phi _\mu =\Omega _\phi A_\mu \Omega ^\dagger
_\phi +ig^{-1} \Omega _\phi \pl _\mu \Omega ^\dagger_\phi )\ .
\label{20}
\end{equation}
Finally, we split $A^\phi _\mu $ into a sum (\ref{12}) of its diagonal
(Cartan) and off-diagonal components. The residual abelian gauge
transformations assume the form (\ref{14}), while the field
$h$ is invariant. If we were able to calculate
the integral over $h$ (it is not gaussian due to the Faddeev-Popov
determinant associated with the unitary gauge (\ref{19})), 
then we would get a
non-local functional of $A_\mu$ as a pre-exponential factor in (\ref{3})
that would be invariant only with respect to the maximal abelian
group $G_H$. Thus, the condition (\ref{19}) is not a gauge condition
imposed directly on gauge fields, but nonetheless it breaks the
gauge symmetry to the maximal abelian subgroup through the
dynamical coupling of the auxiliary scalar and gauge fields.
For this reason we call it the dynamical abelian projection.

Now we turn to the analysis of singular configurations that can occur in
the dynamical abelian projection (\ref{20}). The group element $\Omega
_\phi$ is ill-defined at spacetime points  where the Jacobian $\mu
(h)$ of the change of variables $\phi =\Omega ^\dagger_\phi h \Omega _\phi $
vanishes. It is not hard to calculate it \cite{lata}. If we denote $d\phi
=\mu (h)dhd\mu _H(\Omega _\phi )$, where $d\mu _H(\Omega _\phi )$ is a
Haar measure on $G/G_H$, then for SU(2) and SU(3) we get respectively
\begin{eqnarray}
\mu (h) &= &\mu (\phi )={\rm tr}\, \phi^2 \ ,\label{21}\\
\mu (h) &= &\mu (\phi )={\textstyle\frac{1}{2}}
({\rm tr}\, \phi ^2)^3-({\rm tr}\, \phi
^3)^2\ .\label{22}
\end{eqnarray}
For an arbitrary group, the Jacobian can be found in \cite{lata}. In the
case of SU(2), the Jacobian vanishes when the scalar field has zeros,
$\phi _a(x)=0$. This imposes three conditions on four spacetime
coordinates and therefore determines a set of world lines; the same
holds for SU(n) \cite{lata}. Thus, $\Omega _\phi $ is ill-defined
on some set of worldlines that are formed by zeros of a gauge invariant
polynom $\mu(\phi)$.
To show that these worldlines are worldlines of
magnetic monopoles, one has to repeat the analysis
(\ref{14a})--(\ref{14f}) of section 2, replacing
the operator $\Gamma$ by $\phi$. 

We shall not do this, instead we give an example of a monopole
located at the origin $\mb{x}_m=0$. The field $\phi _a$ must
vanish at $\mb{x}_m=0$, so we choose $\phi _a=x_af (\mb{x})$ (time is
fixed), $ f(0)={\rm const}$. It is easy to find a matrix $\Omega _\phi \in
SU(2)$ that diagonalizes the $2\times 2$ matrix $\phi =\phi _a\tau _a$.
We obtain for (\ref{14e})
\begin{equation}
\bar{C}_\mu^{\phi} =
g_m\left(1+{x_3}/{r}\right)\pl _\mu \tan ^{-1}({x_2}/{x_1})
\ ,\ \ \ \ \ r=|\mb{x}|\ ,\ \ \ \ g_m = (2g)^{-1}
\label{29}
\end{equation}
which is the vector potential of the Dirac monopole with the magnetic
charge $g_m$. 
The curl of the monopole vector potential can be split into a sum of
the Coulomb field 
\begin{equation}
\bar{\mb{B}}_{coul}=g_m\mb{\nabla}r^{-1}\ ,
\label{30}
\end{equation}
and the string field 
\begin{equation}
\bar{\mb{B}}_{st}=g_m\mb{\eta}_3\left(1+{x_3}/{r}\right)[\pl _1,
\pl _2]\varphi(x_1,x_2)\ ,
\label{31}
\end{equation}
where $\varphi (x_1,x_2)$ is the polar angle on the $x_1x_2$-plane and
$\mb{\eta}_3$ is the unit vector pointing in the $x_3$-direction. Clearly,
the Dirac string is located at the semiaxis $x_1=x_2=0$ and $x_3> 0$.
The two terms in the flux integral (\ref{14dd}) correspond to
contributions
of (\ref{30}) and (\ref{31}), respectively. 
The string field (\ref{31}) is
not observable, meaning that it should not contribute 
to the monopole magnetic field energy.
In the next section we give an explicit construction of configurations
that do not give rise to observable Dirac strings upon the dynamical
projection. 

Thus, all topological
singularities occurring upon the dynamical abelian projection are
magnetic monopoles; their charges are given by the topological number
(\ref{14c}) times the inverse coupling constant $(2g)^{-1}$,
and their locations are determined by zeros of the gauge
invariant polynom $\mu (\phi )$ (cf. (\ref{21}) and (\ref{22})). The
auxiliary scalar field $\phi$ can be viewed as the "monopole" field that
carries all information about monopoles in the dynamical abelian
projection. 

To make a use of the dynamical abelian projection in the continuum
limit, we need to establish a fact of the monopole dominance in this 
projection. We now proceed to prove it and give an explicit
parametrization of the dominant configurations via the 
auxiliary scalar filed $\phi$.

\subsubsection*{5. Universality of the dynamical abelian projection}

As has been pointed out in the previous section, the magnetic field
of the Dirac string is not observable. To remove a contribution
of the Dirac string to the magnetic field energy, we have to seek
for such configurations of gauge fields that the associated field
strength $F_{\mu\nu}$ does not exhibit any string-like singularity
after the projection. With this purpose, consider a configuration
\begin{equation}
\bar{A}_\mu(\phi) = (4ig)^{-1} [e_\phi, \pl_\mu e_\phi]\ .
\label{m1}
\end{equation}
Its characteristic property is
\begin{equation}
D_\mu (\bar{A}(\phi)) e_\phi = 0\ .
\label{m2}
\end{equation} 
Since $[D_\mu,D_\nu] = (ig)^{-1}F_{\mu\nu}$ we conclude that
$[F_{\mu\nu},e_\phi]=0$ and, hence,
\begin{equation}
F_{\mu\nu} = e_\phi G_{\mu\nu}^\phi\ ,
\label{m3}
\end{equation}
where $G_{\mu\nu}^\phi$ is a gauge invariant tensor. A simple 
calculation yields
\begin{equation}
G_{\mu\nu}^\phi 
= {\textstyle \frac 12}{\rm tr}\, e_\phi F_{\mu\nu}=
= (8ig)^{-1}{\rm tr}\, \left(e_\phi 
[\pl_\mu e_\phi,\pl_\nu e_\phi]\right)\ .
\label{m4}
\end{equation}
Upon the projection $e_\phi$ goes over to $\tau_3$ and, hence,
$F_{\mu\nu}$ becomes purely abelian. According to (\ref{m4})
it has no string-like singularities. The gauge potential
(\ref{m1}) also turns into a purely abelian one 
\begin{equation}
\bar{A}_\mu(\phi)\rightarrow \tau_3 (C_\mu^{D} + C_\mu^{st})\ ,
\label{m5}
\end{equation}
where $C_\mu^D$ is the monopole  potential, and $C_\mu^{st}$
has a support on the Dirac string; its curl produces the 
string magnetic field that cancels with the string field 
coming from the curl of the Dirac potential. In a particular
case when $\phi_a\sim x_a$, $\mb{C}^{st}$ is constant along
the half-line $\theta =0$ (here $\cos\theta = x_3/r$), whereas
its curl is proportional to $\delta(x_1)\delta(x_2)\theta_H(x_3)$
with $\theta_H$ being the Heaviside function.

Thus, the configuration (\ref{m1}) has a desired property:
upon the projection the associated abelian field strength 
does not have the singularity along the Dirac string.

We define a conservative monopole current as
\begin{equation}
j_\mu^m = (8\pi)^{-1} \varepsilon_{\mu\nu\sigma\tau}
\pl_\nu G_{\sigma\tau}^\phi \ ,\ \ \ \ \pl_\mu j^m_{\mu} =0\ .
\label{m6}
\end{equation}
The magnetic charge density $j_0^m$ is then determined by
$(\mb{\nabla},\mb{B})$ as follows from (\ref{m4}) and
(\ref{14c}).

The magnetic field energy of  monopoles is still divergent 
because monopoles are point-like particles. To regularize it,
we have to give the Dirac monopole a core as prescribed by
(\ref{dc}). It can be achieved, for instance, by the replacement
\begin{equation}
\bar{A}_\mu (\phi )\rightarrow \bar{A}_\mu(\phi) +
\bar{A}^+_\mu (x) e^+_\phi +
\bar{A}^-_\mu (x) e^-_\phi\ ,
\label{33}
\end{equation}
where
$e^\pm _\phi =\Omega ^\dagger_\phi \tau _\pm \Omega _\phi\ ,
\tau _\pm =\tau _1\pm i\tau _2$
and $\bar{A}^\pm _\mu $ are the core functions
\cite{kogut},
that is, upon the projection 
the configuration (\ref{33}) gives rise to a Wu-Yang monopole
with  a core \cite{kogut}. In the conclusion we propose
an alternative regularization.

Let us now turn to the question of the dominance of the configurations
(\ref{33}).
By construction, the subspace $[\bar{A}(\phi)]$ of all such
configurations is one of the "monopole" subspaces $[ ^\chi \bar{A}]$
obtained via an abelian projection in section 3.
Positions of Dirac monopoles in the dynamical abelian projection
are determined by zeros of the gauge invariant polynom $\mu(\phi)$.
Since the auxiliary field $\phi$ fluctuates, the Jacobian $\mu(\phi)$
fluctuates too, so do the positions and charges of the monopoles.
Therefore, letting all configurations of $\phi$ to occur, which is
indeed the case according to the identity (\ref{17}), we can generate
{\em all} possible distributions of monopoles (\ref{14c}) by making the
dynamical abelian projection. Hence, the set of monopole
distributions $[q_\phi]$ in the dynamical abelian projection is not
less than the projection independent set (\ref{dc3})
\begin{equation}
[\bar{q}] \subseteq [q_\phi]\ .
\label{md}
\end{equation}
On the other
hand, we have established a one-to-one correspondence between
monopole distributions from the set (\ref{dc3}) and the subspace
of the dominant configurations $[\bar{A}]$ modulo regular gauge
transformations (see (\ref{dc3})). From this observation and
the relation (\ref{md}) follows that the subspace $[\bar{A}(\phi)]$
covers the subspace of the dominant configurations  $[\bar{A}]$
\begin{equation}
[\bar{A}]/[\Omega_0] \subseteq [\bar{A}(\phi)]/[\Omega_0]\ .
\label{md2}
\end{equation}
Therefore any configuration from the dominant subspace $[\bar{A}]$
can be represented in the form (\ref{33}).

One can understand this from another point of view. By definition
every configuration $\bar{A}_\mu$ turns into a Dirac monopole with
a core after a certain abelian projection $\chi$. In turn the abelian
projection implies a gauge rotation of $\bar{A}_\mu$ with a singular
group element $\Omega_\chi$ specified  by $\chi$. The relation
(\ref{md2}) follows from two facts. First, it is
always possible to find such a configuration of the auxiliary
field $\phi(x)$ that the group element $\Omega_\phi$ would lead
to the very same monopole distribution (\ref{14c}) as does the
$\Omega_\chi$, and, second, the
configuration (\ref{33}) turns into the Dirac monopole
with a core after the gauge rotation (or projection) with
the group element $\Omega_\phi$.

Thus, the dominant configurations are parametrized by an auxiliary
scalar field $\phi$ in the adjoint representation, $\bar{A}_\mu
=\bar{A}_\mu(\phi)$, that is, the field $\phi$ plays the role
of the {\em field collective} coordinate for the dominant configurations.
The trick of inserting the identity (\ref{17}) into the path
integral (\ref{3}) can be regarded as a way to obtain a right
{\em gauge} and {\em Lorentz} invariant measure for the collective
coordinate. Note that the dominant configurations are determined
up to regular gauge transformations, therefore any parametrization
of the dominant subspace should be invariant under regular gauge
transformations. The latter does hold for our parametrization
because the monopole distribution (\ref{14c}) is determined by
zeros of the {\em gauge invariant} polynom $\mu(\phi)$ and
the measure for the collective field coordinate  in (\ref{17})
respects the gauge symmetry too.

\subsubsection*{6. An effective theory for dominant configurations}

To develop an effective theory for the dominant configurations,
in the integral
\begin{equation}
\< W_R\>=\< 1\>^{-1}\int {\cal D}A_\mu {\cal D}\phi 
\sqrt{\det(-D^2_\mu)}
e^{-S_{YM}-\frac 12\int d^4x{\rm tr}(D_\mu\phi)^2}W_{R}(A)\ ,
\label{36}
\end{equation}
we perform the change of variables
\begin{equation}
A_\mu=\bar{A}_\mu (\phi)+a_\mu\ ,
\label{37}
\end{equation}
and make the gaussian approximation for quantum fluctuations $a_\mu$ around
the configurations (\ref{33}). In the new integration variables
$(\phi ,a_\mu)$ the measure has the standard form ${\cal D}A_\mu {\cal
D}\phi={\cal D}a_\mu {\cal D}\phi$ because the new potentials $a_\mu$ are
obtained by a shift of $A_\mu$ on a function of the
independent integration variable $\phi$. It should be emphasized that
this procedure is no semiclassical approximation because $\bar{A}_\mu(\phi)$
are fully quantum (no classical) configurations of gauge fields parametrized
by the quantum field $\phi$ as prescribed by (\ref{33}).

To regularize a divergence of the path integral (\ref{36}) 
caused by  the
gauge symmetry, it is convenient to use the background gauge
\begin{equation}
\pl _\mu A_\mu +ig[\bar{A}_\mu (\phi ), A_\mu ]=0\ .
\label{38}
\end{equation}
So, the corresponding Faddeev-Popov determinant is to be inserted into the
path integral measure in (\ref{36}).
Provided the monopole dominance holds in the continuum
limit (which is
supported by the  lattice simulations \cite{stack,bali}),
the background gauge
is safe regarding the Gribov problem because the integral over
$A_\mu$ in (\ref{36})
is dominated by
the background configurations (\ref{33}), and,
therefore, can be calculated by the perturbation theory over $a_\mu$.
Doing the Gaussian integral over $a_\mu$, one gets
\begin{equation}
\< W_R\> \approx \< 1\>^{-1}
\int {\cal D}\phi e^{-S_{eff}(\phi, J_0)}
W_{R}(\bar{A}(\phi))
\label{39}
\end{equation}
for the Wilson loop expectation value. The effective action
$S_{eff}$ consists of two parts: the Yang-Mills action of
the background monopole configuration (which is nothing but
the dual QED action) and the  ``trace-logs'' of three functional
determinants taken on the background configuration, which
determine the entropy of the topological (monopole) defects. 
The integral (\ref{39})
describes a quantum theory of the dominant
gauge field configurations, that give rise to Dirac monopoles
upon the dynamical abelian projection, as
an effective 
dynamics of the
auxiliary quantum field $\phi$.

The integration over $a_\mu$ is nothing but an average over small
fluctuations of charged and photon fields in abelian projections.
Therefore the dependence of the effective action $S_{eff}$ on the
external static source $J_0$ involves only the Coulomb interaction
and possible radiative corrections to it (in higher orders of the 
perturbation theory). The integral (\ref{39}) determines an average
of the Wilson loop over the ``monopole'' (dominant) configurations.
Note that if one restricts the integration domain by configurations
of $\phi$ for which $\mu(\phi)\neq 0$, then 
the configuration (\ref{m1}) is a pure gauge (no defects)
and the core functions vanish so that 
the ordinary Faddeev-Popov perturbation
theory is recovered.  

There is a natural generalization of our construction to an arbitrary
unitary group. Let
$\lambda^p,
\lambda^\alpha$
be respectively 
basises in the non-Cartan and Cartan subspace of a Lie algebra. 
The dominant configurations can be parametrized as
\begin{equation}
\bar{A}_\mu (\phi)
= (4ig)^{-1}[e_\phi^\alpha,\pl_\mu e_\phi^\alpha] + 
\bar{A}_\mu^p(x) e_\phi^p \ ,
\label{gen}
\end{equation}
where $\bar{A}_\mu^p(x)$ are core functions and 
$e_\phi^{p,\alpha} = \Omega_\phi^\dagger \lambda^{p,\alpha}
\Omega_\phi,\ \Omega_\phi\in G/G_H$. Quark fields can also be
included.

\subsubsection*{7. Conclusion}

Based on the results of numerical simulations in lattice gauge
theories,
we have proposed a gauge invariant formulation of the monopole
dominance.
Our approach relies on the dynamical abelian projection which
is manifestly Lorentz invariant and exempt of the Gribov problem.
Assuming that the result of numerical simulations, 
that the monopole dominance occurs in at least one abelian
projection, holds in the continuum limit, we have proved
the monopole dominance for the dynamical abelian projection.
The latter enables us to obtain an effective theory
for the dominant configurations in pure Yang-Mills theory.  
A further study of the effective theory will be given elsewhere.

As a final remark we mention also that there exists an abelian
projection of the theory (\ref{36}) that interpolates the maximal
and dynamical abelian projections:
\begin{equation}
\phi^{off} - i\kappa [\chi_{ma}(A), \phi^H]=0\ ,
\label{kappa}
\end{equation}
where the Lie-algebra-valued function $\chi_{ma}(A)$ is defined
by the first equality in (\ref{15}), $\phi^{off,H}$ are off-diagonal
and Cartan components of $\phi$, respectively. A constant $\kappa$
plays the role of the interpolating parameter. The maximal abelian
projection is reached in the limit $\kappa\rightarrow \infty$, while
the dynamical abelian projection corresponds to $\kappa =0$. Applying
arguments similar to those in \cite{thooft2}, one can convince oneself
that the mass parameter $\kappa^{1/2}$ provides a regularization
of topological defects occurring in the $\kappa$-projection (\ref{kappa}).
This could be used as an alternative regularization of the monopole
effective action.

\newpage
\begin {center}
{\bf Acknowledgment}
\end{center}
I wish to thank  M. Asorey,
A. Hart, H. Markum, J. Mourao, M. Polikarpov and F. Scholtz 
for stimulating and fruitful discussions and
their interest in this work. I am also grateful to the
Department of Theoretical Physics of the Valencia University
for the warm hospitality during the time when the main part
of this work has been done.

\end{document}